\documentclass[]{AO4ELT}  %>>> use for US letter paper
\usepackage{microtype}
\usepackage{biblatex}
\usepackage{amsmath,amsfonts,amssymb}
\usepackage{graphicx}
\usepackage{pst-all} % Pour pstricks
\usepackage[colorlinks=true, allcolors=blue]{hyperref}
\makeatletter         
\def\@maketitle{
\includegraphics[width = 170mm]{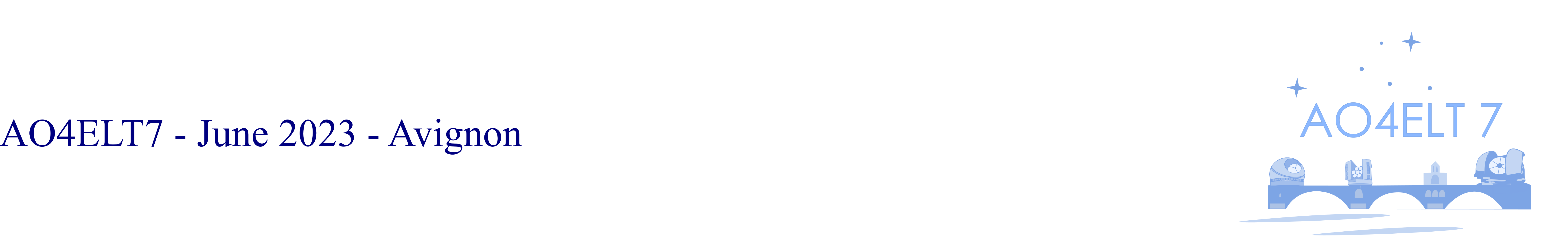}\\[8ex]
\begin{center}
{\Huge \bfseries \sffamily \@title }\\[4ex] 
{\Large  \@author}\\[4ex] 
\@date%\\[8ex]
\end{center}}
\title{Sequential coronagraphic low-order wavefront control}

\author[a]{Michael Bottom}
\author[a*]{Samuel A. U. Walker}
\author[a*]{Ian Cunnyngham}
\author[b]{Charlotte Guthery}
\author[b]{Jacques-Robert Delorme}
\affil[a]{Institute for Astronomy, University of Hawai'i, 640 N. Aohoku Place, Hilo, HI, 96720, United States}
\affil[b]{W. M. Keck Observatory, 65-1120 Mamalahoa Hwy, Waimea, HI 96743, United States}

\authorinfo{Send correspondence to MB, mbottom@hawaii.edu\\ *contributed equally to this work}

\begin{document} 
\maketitle
\begin{abstract}
Coronagraphs are highly sensitive to wavefront errors, with performance degrading rapidly in the presence of low-order aberrations.  Correcting these aberrations at the coronagraphic focal plane is key to optimal performance.  We present two new methods based on the sequential phase diversity approach of the ``Fast and Furious'' algorithm that can correct low-order aberrations through a coronagraph.  The first, called ``2 Fast 2 Furious,'' is an extension of Fast and Furious to all coronagraphs with even symmetry.  The second, ``Tokyo Drift,'' uses a deep learning approach and works with general coronagraphic systems, including those with complex phase masks.  Both algorithms have 100\% science uptime and require effectively no diversity frames or additional hardware beyond the deformable mirror and science camera, making them suitable for many high contrast imaging systems.  We present theory, simulations, and preliminary lab results demonstrating their performance.
\end{abstract}


\section{INTRODUCTION}
\label{sec:intro}  % \label{} allows reference to this section

Coronagraph designs have advanced greatly in recent years, driven by the needs of ``extreme'' adaptive optics systems (ExAO, \cite{guyon_extreme_2018}) operating on large ground-based telescopes (SCExAO, GPI, SPHERE, MagAO-X, etc).  Modern coronagraphs like the PIAA \cite{guyon_phase-induced_2003} or vortex \cite{mawet_annular_2005} deliver close to optimal performance within the limits imposed by wave optics. However, there are fundamental trades between coronagraph contrast and inner working angle, and sensitivity to low-order aberrations such as tip/tilt, focus, astigmatism, etc.  Even in coronagraphs with modest performance, when aiming for high contrast, the sensitivity to low-order aberrations is important. 

\begin{figure}
    \centering
    \includegraphics[scale=0.3]{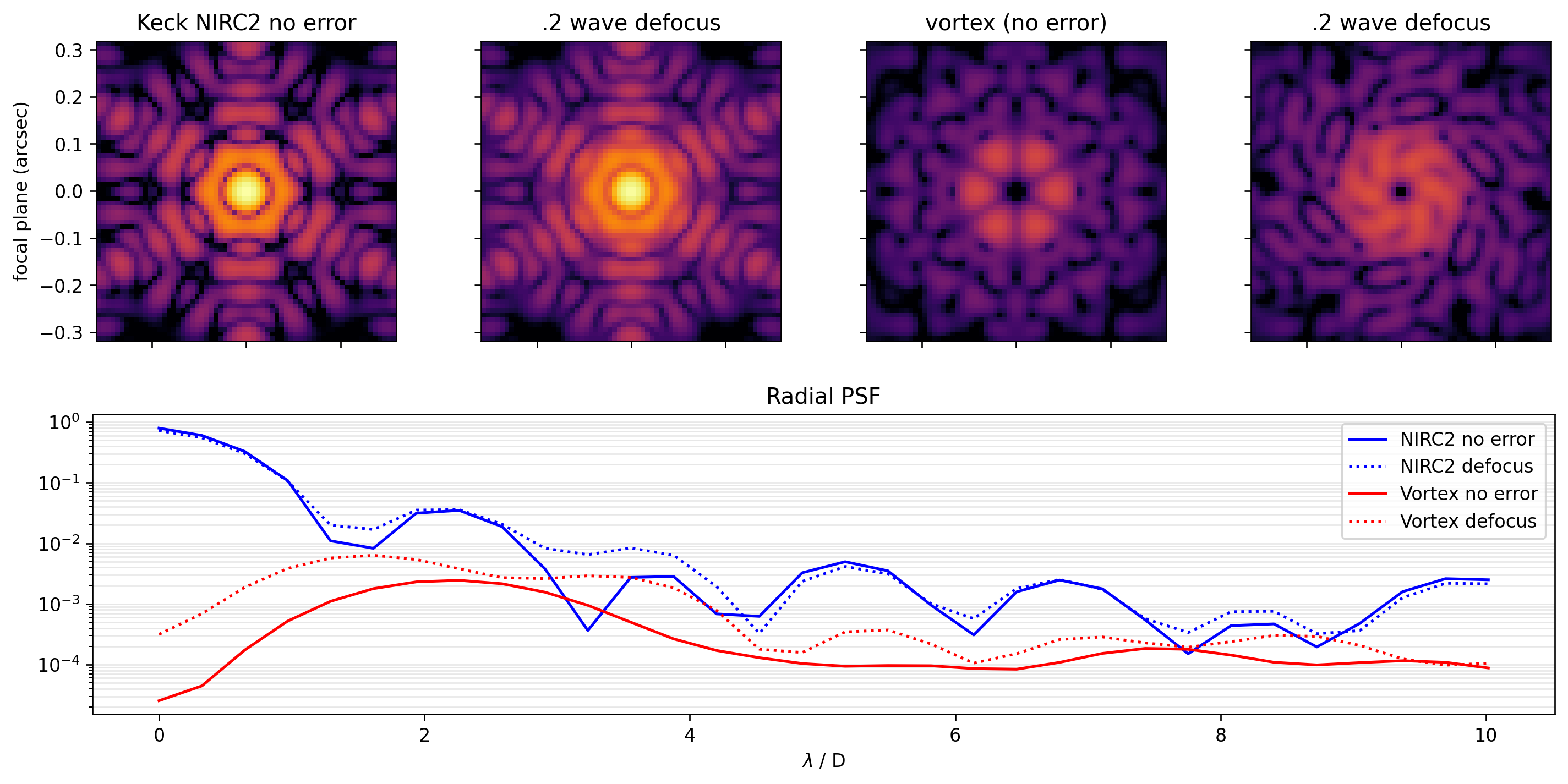}
    \caption{Simulation of Keck's NIRC2 PSF with no errors, with .2 (wave / D) defocus, then repeated with a vortex coronagraph, demonstrating the much higher sensitivity of the vortex to small aberrations}
    \label{fig:vortex_defocus}
\end{figure}

Low-order aberrations are not all equal, however.  Aberrations occurring before the coronagraphic focal plane are more destructive than those occurring after.  For example, Figure \ref{fig:vortex_defocus} shows a small (0.2 $\lambda$) defocus term applied to a unocculted PSF (such as a planet), which barely affects the peak brightness (eg, Strehl ratio).  However applying the same error before a coronagraphic focal plane (in this case a vortex) leads to degradation in contrast by almost an order of magnitude.

Given this sensitivity to low-order aberrations, there are two complementary approaches to sensing and controlling these aberrations at the coronagraph plane.

One way to solve such a problem is to use extra hardware to detect reflected or refracted light from the coronagraph optics and use this to determine the level of aberration in the system.  Guyon 2009 \cite{guyon_coronagraphic_2009} proposed such a scheme using a defocused image of light reflected from a coronagraphic spot.  The high contrast imaging system SPHERE on the VLT and the Stellar Double Coronagraph at Palomar used similar approaches to control pointing only on the coronagraph, with a tip/tilt sensor hardware mounted very close to the coronagraphic focal plane \cite{baudoz_differential_2010, bottom_stellar_2016}.  Beyond tip and tilt, Singh et al. 2014 \cite{singh_lyot-based_2014} demonstrated control of many low-order modes using light reflected from Lyot-style coronagraphic masks.  In these proceedings Haffert et al. also presents a near-optimal wavefront sensor based on beam apodizing optics and a Zernike phase mask.  The main drawbacks of hardware-based coronagraphic low-order wavefront sensors is that they require extra hardware. 

An alternative method is to use the the science camera focal plane as the wavefront sensor, imaging the light transmitted through the coronagraph.  This approach is similar to PSF sharpening algorithms in non-coronagraphic imaging.  For example, The Gerchberg-Saxton algorithm uses known injected phase diversity (usually defocus) to solve for the residual phase aberrations in the pupil plane, which can then be corrected resulting in Strehl improvement.  The need for the phase diversity is derived from the well-known even sign degeneracy from symmetric pupils such as those on most telescopes.  Asymmetric pupils allow for direct sensing of all modes \cite{martinache_asymmetric_2013} without phase diversity, as can other specialized (asymmetric) optical designs like vector apodizing phase plates \cite{bos_focal-plane_2019} or self-coherent cameras \cite{baudoz_self-coherent_2005}.  These may be seen as incorporating wavefront sensing directly into the coronagraph design, but their principles and operation are not applicable to general coronagraphic systems.  

Focal-plane wavefront control through a general coronagraphic system was first described in Paul et al. with the COFFEE algorithm \cite{paul_high-order_2013}, which used a phase-diversity approach (using large defocus and astigmatism terms) to derive aberrations upstream and downstream of the coronagraph.  This approach was shown to work well in simulation and was demonstrated using the internal source at SPHERE \cite{paul_compensation_2014}. While it can be used with any coronagraphic system, it has the disadvantage of requiring two diversity frames per correction step, reducing science time.  

Ideally, the focal plane wavefront control algorithm will require minimum or no diversity frames, and like COFFEE, work with any coronagraph.  Additionally, it should work reliably in an operational environment, and converge rapidly.  For example, an algorithm called QACITS \cite{huby_qacits_2016} to control just tip/tilt has been in regular on-sky use with the Keck observatory vortex coronagraph \cite{serabyn_w_2017, huby_-sky_2017} effectively correcting slow pointing drifts off the vortex and dramatically improving performance.  At its core, QACITS uses focal-plane images of the tip/tilt response to build an  linearized model.  While QACITS was developed specifically for a vortex coronagraph, the main ideas could be applied to almost any coronagraphic system to control tip/tilt.  Orban de Xivry et al. \cite{orban_de_xivry_post-coronagraphic_2017} extended this work to derive many low-order aberrations through a vortex coronagraph using the weak phase solution (ie small aberrations applied), and with two phase diversities  applied.  This technique was shown to work very well in simulation, but relied on an analytic formulation only applicable to a vortex coronagraph with an unobstructed pupil.

In summary, we submit that an ideal low-order focal-plane wavefront control algorithm would possess the following properties:
\begin{enumerate}
\item Work reliably on-sky and converge quickly
\vspace{-2.5mm}\item Require no extra hardware beyond a science camera and deformable mirror
\vspace{-2.5mm}\item Require no diversity frames
\vspace{-2.5mm}\item Work with arbitrary pupils
\vspace{-2.5mm}\item Work with any coronagraph
\end{enumerate}

It would seem that points 2, 3, and 4 are in mutual contradiction, as a symmetric pupil would naturally inherit the even phase degeneracy ultimately due to Fourier symmetries, and one would need to use extra hardware or diversity frames to determine the sign.  However, there is a practical workaround to this, which is used in the (non-coronagraphic) sequential phase diversity algorithm Fast and Furious.

Fast and Furious is a high-performance focal-plane PSF optimization algorithm.  First developed in 2013 by Keller et al. \cite{keller_extremely_2012}, it is based on the sequential phase-diversity approach of Gonsalves \cite{gonsalves1982phase} in the weak-aberration limit. An on-sky astronomical demonstration occurred at the Subaru telescope in 2019, generating excitement by showing potential for correcting the notorious ``low-wind effect'' \cite{bos_-sky_2020}, where thermal discontinuities create large piston errors between telescope spiders, destroying PSF quality.  Fast and Furious uses the following steps to sense the aberrations:

\begin{enumerate}
    \item The PSF image is divided into even (symmetric) and odd (antisymmetric) parts. 
    \vspace{-2.5mm}\item The odd part can directly solve for the odd electric field. 
    \vspace{-2.5mm}\item The even part can find the amplitude of the even electric field but not its sign. 
    \vspace{-2.5mm}\item Using the previous PSF and correction, the sign of the even component is determined.
    \vspace{-2.5mm}\item The sign and amplitude components provide the full even electric field. 
    \vspace{-2.5mm}\item By Fourier transforming the odd and even electric field, the complete wavefront error is reconstructed. 
    \vspace{-2.5mm}\item A correction is made ($\Phi_i \rightarrow \Phi_d$), and the process repeats.
\end{enumerate}
\begin{figure}
    \centering
    \includegraphics[width=0.75\textwidth]{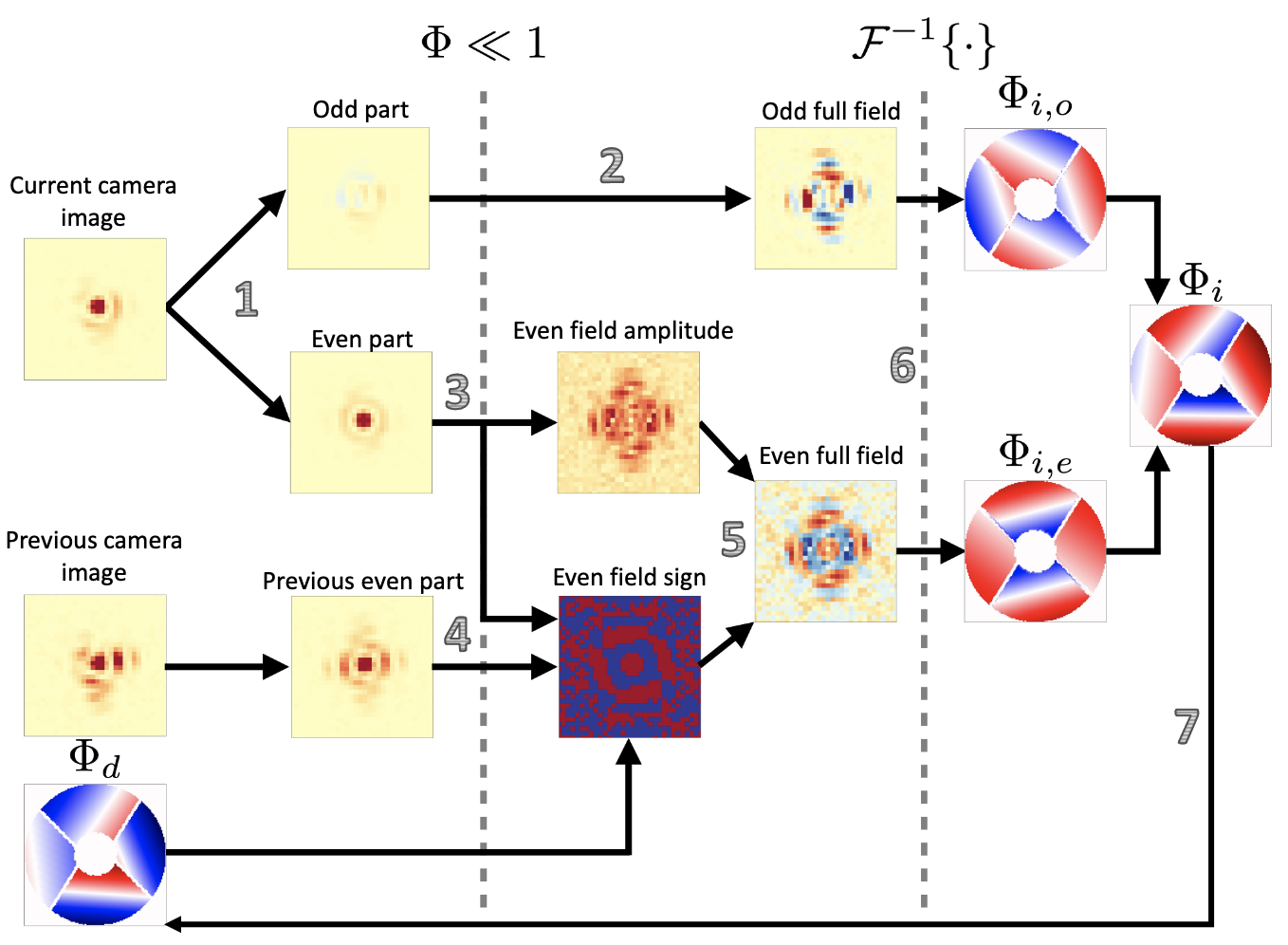}
    \caption{Schematic of the Fast and Furious algorithm.  Refer to the text for details.  Schematic adapted from Bos et al. 2021}
    \label{fig:enter-label}
\end{figure}

Fast and Furious has also been used on-sky at Keck on many occasions (\textcite{bos_fast_2021} also see Guthery et al., these proceedings), and has demonstrated reliable and rapid convergence.  One key idea of the algorithm is to use the previous correction as the diversity phase, which means the PSFs stays very close to optimal at every iteration, and no science time is lost to large diversity phases.  Indeed, the algorithm can be run concurrently with observations, which also tends to stabilize the PSFs, leading to better systematics control.  The other aspect which is helpful is that it only requires a model of the pupil and uses Fraunhofer progagators, so calculation speed and complexity is minimal.  In its native implementation, F\&F has several requirements which preclude operation with arbitrary optical systems: the exit pupil must be real and symmetric, the PSF must be distorted only by phase aberrations, and the PSF must be shift-invariant.  The only coronagraph design that fulfils these three conditions is a symmetric apodized pupil coronagraph.

Therefore, the goal of this work is to try to modify F\&F to work with arbitrary coronagraphic systems.  The first method we present, nominally called ``2 Fast 2 Furious,'' (2F2F) is a linear method  similar to the regular F\&F algorithm but including some extra coronagraphic operators.  2F2F nearly fulfils the ideal algorithm criteria presented in the list above, but is restricted to work on symmetric coronagraphs, such as classic Lyots, PIAAs, or several classes of shaped pupil masks.  It does not work on hybrid designs or with complex phase masks like the vortex coronagraph.  The second method, informally referred to as ``Tokyo Drift,'' (TD) is F\&F in spirit only, with the same trick of using the previous correction to break the sign degeneracy, but uses a neural network trained on the optical model to derive the aberrations.  The benefit of all this extra work is flexibility, as TD can work with any type of coronagraph or optical system, and can correct well outside the weak aberration regime ($>>$1 radian rms phase error).

With respect to Tokyo Drift, relevant earlier work may be found in Quesnel \cite{quesnel_deep_2020} and Orban de Xivry \cite{orbandexivry_focal_2021}.  These papers demonstrate the power of deep learning for focal-plane wavefront sensing (and with coronagraphs), including measuring the performance compared with fundamental limits.  However, these works use either phase diversity or specific optical designs allowing for even-mode sign disambiguation, such as scalar vortex coronagraphs or polarization optics.  The primary difference in our approach is that we present a general method--based on Fast and Furious--that can work for any coronagraph design.  For both 2F2F and TD, we also provide preliminary experimental demonstration.  For TD, this demonstrates that despite being trained on simulated data, the network still perform well on real hardware.

In the following sections, we present more detailed discussion of these approaches, results from simulations, and AO bench results from Keck (for 2 Fast 2 Furious) and Subaru (Tokyo Drift).

\section{2 Fast 2 Furious}

\subsection{Mathematical formalism}

2 Fast 2 Furious is an extension of the mathematical formalism of Fast and Furious designed to work with symmetric amplitude-mask coronagraphs such as the Lyot coronagraph. In brief, we insert extra terms in to the master equation that \cite{korkiakoski_fast_2014} use to describe the propagation of light from pupil to focal plane which represent the coronagraphic focal plane mask and Lyot stop. We find that the methodology of \cite{korkiakoski_fast_2014} is still relevant in this regime, providing some substitutions are made. Our calculations are presented below.

The Fast and Furious algorithm is based upon an expansion of the equation 

\begin{equation}
    p = |\mathcal{F}\{A~\textrm{exp}(i\phi)\}|^2,
    \label{eq:propnocoro}
\end{equation}
which describes the final focal plane image $p$ as a Fourier transform ($\mathcal{F}$) of an electric field with amplitude $A$ and phase $\phi$ from the pupil plane (where the incoming wavefront is defined) to the focal plane (where the observed image is defined). The case of $\phi=0$, corresponds to an unaberrated image with no phase error. When using a coronagraph, one typically places a focal plane mask (denoted by $C$) and a pupil plane Lyot stop (denoted by $M$) into this optical path, thus:

\begin{equation}
    p = |\mathcal{F}\{M\mathcal{F}^{-1}\{C\mathcal{F}\{A~\textrm{exp}(i\phi)\}\}\}|^2.
    \label{eq:propcoro}
\end{equation}
Here, $M$, $C$ and $A$ are all properties of the telescope system in use, as $A$ is defined by the telescope aperture. One defining assumption of the Fast and Furious approach is that the aperture $A$ is a real and even function of pupil plane position. Without this assumption the expansion out of the modulus in equation \ref{eq:propnocoro} involves several terms that would otherwise cancel, significantly complicating the math. If both $M$ and $C$ are also real and even functions, as is the case for many coronagraphs, then the three system properties in equation \ref{eq:propcoro} can be combined into one real and even system parameter, hereinafter denoted by $X$ and given by

\begin{equation}
    X = M\mathcal{F}^{-1}\{C\mathcal{F}\{A\}\}.
    \label{eq:x}
\end{equation}
Substituting this back into equation~\ref{eq:propcoro} gives that the image is described by the equation

\begin{equation}
    p = |\mathcal{F}\{X~\textrm{exp}(i\phi)\}|^2,
    \label{eq:propcorowithx}
\end{equation}
which is entirely analogous to equation \ref{eq:propnocoro} with the substitution $A \rightarrow X$. 

This methodology is summarised below. We adopt the notation that the Fourier transform of a quantity denoted by an upper-case letter is denoted by a lower-case letter, i.e. $x = \mathcal{F}\{X\}$.

Firstly, the exponential in equation \ref{eq:propcorowithx} is expanded to first order in $\phi$, and then the phase is expressed as a sum of its even and odd terms, $\phi = \phi_e + \phi_o$. These terms are carried through to obtain expressions for the even and odd parts of the final image. The odd part of the final image is then given by

\begin{equation}
    p_o = 2ix\mathcal{F}\{X\phi_o\},
    \label{eq:po}\\
\end{equation}
where $i = \sqrt{-1}$. Similarly, the even part of the final image is given by

\begin{equation}
    p_e = x^2 + \left(\mathcal{F}\{X\phi_e\}\right)^2 - \left(\mathcal{F}\{X\phi_o\}\right)^2,
    \label{eq:pe}
\end{equation}
where it can be seen that if $\phi_o$ is already known from equation \ref{eq:po}, the only unknown becomes $\phi_e$. The F\&F approach breaks the even aberration sign degeneracy evident in this equation by using the deformable mirror correction calculated in the previous iteration as a diversity phase. If $p_n$ is the image at the most recent iteration, and $p_{n-1}$ is the image at the iteration prior to that with a phase offset of $\phi_d$ from $p_n$ (given by the deformable mirror correction applied from the previous iteration's calculations), then the difference between the even parts of these two images can be written as

\begin{equation}
\begin{split}
    p_{e, n} - p_{e, n-1} & = \left(\mathcal{F}\{X\phi_{o,d}\}\right)^2 - \left(\mathcal{F}\{X\phi_{e,d}\}\right)^2 \\
    & + 2\mathcal{F}\{X\phi_o\}\mathcal{F}\{X\phi_{o,d}\} - 2\mathcal{F}\{X\phi_e\}\mathcal{F}\{X\phi_{e,d}\}.
    \label{eq:diffp}
\end{split}
\end{equation}

\cite{korkiakoski_fast_2014} find their version of this equation to be unstable due to the subtraction of two similar images, using it purely to find the sign of $\phi_e$ in combination with the absolute value of the $\phi_e$ obtained from equation \ref{eq:pe}. We continue to follow this approach going forward.

\subsection{Testing and results}

Despite the mathematical substitutions above, it was not clear that algorithm would work in practice due to the violation of conservation of energy in coronagraphic systems from pupil to focal plane, an implicit assumption in Fast and Furious \cite{keller_extremely_2012}.  We implemented the algorithm coupled with simulations of the Keck/NIRC2 optical path using the Python package \texttt{HCIPy} \cite{HCIPy}. In contrast to Fast and Furious, we find that 2 Fast 2 Furious performs poorly under direct inversion of equations \eqref{eq:po}, \eqref{eq:pe} and \eqref{eq:diffp}, necessitating a numerical approach for solving these equations. To increase computation speed, we fit for the coefficients of the Zernike modes that we are aiming to correct, which we find works well and performs at speeds on the order of 2-3 seconds, which is efficient enough for eventual on-sky deployment. Throughout our analysis we focus on correcting for the first 20 Zernike modes, under the assumption that the majority of the power of any on-sky aberrations would be in these modes.

\begin{figure}
    \centering
    \includegraphics[keepaspectratio, width=0.5\linewidth]{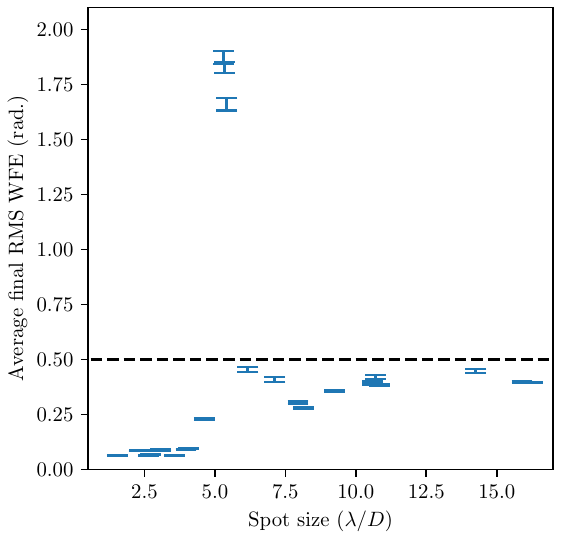}
    \caption{Average final RMS WFE as a function of focal plane mask size in units of $\lambda/D$ for various combinations of focal plane mask and filter wavelength available on Keck/NIRC2. The algorithm performs well for coronagraphs with small spot sizes $\lesssim4\lambda/D$, which is expected given that we are only correcting the first 20 Zernike modes. Errorbars represent the standard error on 100 measurements of each datapoint, with differing initial phases for each measurement.}
    \label{fig:lambdaoverd}
\end{figure}

\begin{figure}
    \centering
    \includegraphics[keepaspectratio, width=0.5\linewidth]{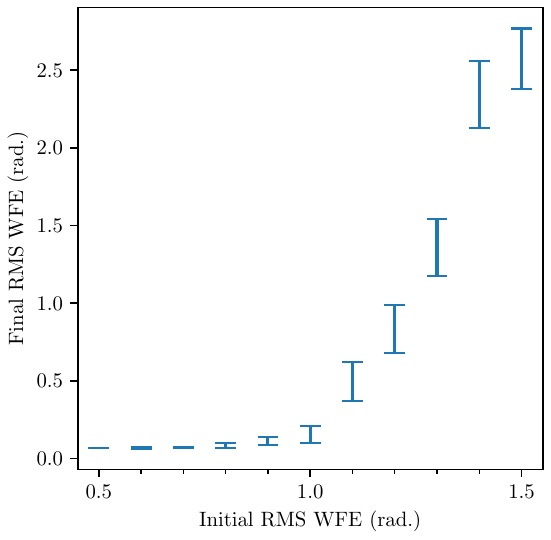}
    \caption{Average final RMS WFE after 20 iterations as a function of initial injected RMS WFE for a 100 mas coronagraph in the $K$-band on Keck/NIRC2. 2 Fast 2 Furious converges reliably every time for initial aberrations with an RMS of less than 1.1 radians. Errorbars represent the standard error on 100 measurements of each datapoint, with differing initial phases for each measurement.}
    \label{fig:rmswfe}
\end{figure}

Using these simulations, we characterised the regime in which 2 Fast 2 Furious converges. The first of our tests focused on the coronagraph sizes. For this, we simulated PSF images with a range of focal plane mask sizes and observing bands based on those currently available for science use on Keck/NIRC2. For each image, we then applied 0.5 radians total root-mean-square wavefront error (RMS WFE) injected randomly across the 20 Zernike modes that we aim to correct, before running the image through 20 iterations of 2 Fast 2 Furious and recording the RMS WFE of the final image. We repeated this process 100 times for each instrument configuration, randomly altering the shape of the wavefront error each time. We found that the algorithm converges best when operated with coronagraphs that have spot sizes $\leq4\lambda/D$, as shown in Figure \ref{fig:lambdaoverd}. Beyond this limit, the phase of the image either diverges from the initial injected value or else converges weakly.  This is likely due to the low signal from Zernikes of index 20 or less at those spatial frequencies.

Our second test concerned the maximum amplitude of aberrations that 2 Fast 2 Furious could reliably correct. We took simulated images of the smallest 100 mas focal plane mask in the $K$-band and performed 100 tests of 2 Fast 2 Furious for various initial input aberrations RMS WFE amplitudes from 0.5 radians to 1.5 radians, again allowing 20 iterations for convergence. This experiment showed that 2 Fast 2 Furious reliably converges when dealing with aberrations below $\sim 1.1$ radians ($1/6$ of a wave), which would correspond to approximately 33\% Strehl ratio without a coronagraph in the optical path (see Fig. \ref{fig:rmswfe}). This value is key as the AO system at Keck is typically able to correct to this level or better \cite{liu2008}, so the fact that 2 Fast 2 Furious can correct images at these Strehl levels demonstrates that the algorithm operates reliably at the point where non-common-path aberrations begin to dominate the wavefront error budget.  It also is the same as the correction range of the non-coronagraphic F\&F algorithm.

\begin{figure}
    \centering
    \includegraphics[keepaspectratio, width=\linewidth]{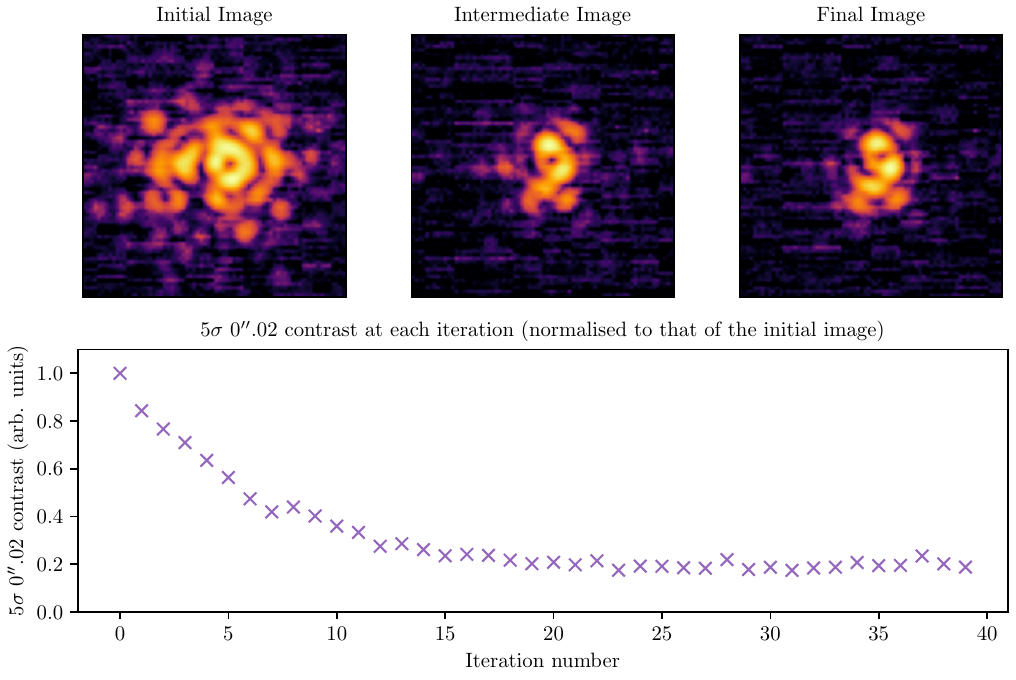}
    \caption{Results from Keck/NIRC2 bench testing when correcting for tip-tilt with the 100mas coronagraph in the optical path. Images are presented in log scale. The $5\sigma~0^{\prime\prime}.1$ contrast improves with each iteration as the PSF becomes sharper and more well-defined, denoting improved PSF/coronagraph alignment.}
    \label{fig:bench}
\end{figure}

Once we had proved that 2 Fast 2 Furious worked in simulations, we sought lab validation. As such, we carried out several tests on the Keck/NIRC2 bench, and these tests are still ongoing. Our main preliminary result is that we have successfully used 2 Fast 2 Furious to correct for injected tip/tilt errors using the 100 mas coronagraph in the $K$-band on the bench. Evidence of this is presented in Figure \ref{fig:bench}, wherein we use the $5\sigma$ contrast at $0.1^{\prime\prime}$ as an indicator of image quality. The significant improvements in limiting contrast, as well as the improvement in PSF quality that is evident from the raw NIRC2 images, has us satisfied that 2 Fast 2 Furious can correct tip/tilt well in a real-world setting. We are now working to extend this to more Zernike modes, to mirror the work we have done in simulation.

\section{Tokyo Drift}

\textcite{korkiakoski_fast_2014} modified the F\&F algorithm to accommodate asymmetric pupils using a Gerchberg-Saxton like approach where the pupil amplitudes are iteratively calculated.  However, this approach was found to have instability issues in the small-phase limit (eg, as the algorithm converged) which caused wildly divergent behavior due to subtraction of similar images leading to ``divide by zero'' errors, requiring further filtering and regularization.  It is likely this issue would be exacerbated with coronagraphs.  We instead opted to investigate a conceptually different approach using deep neural networks.

The deep neural network (DNN) analog of F\&F was trained using supervised techniques where large datasets were simulated comprising PSFs of coronagraphic systems and the random wavefront errors which generated them.  The network then learned to directly estimate the Zernikes that would best correct the observed PSF aberrations via regression.  In operation, this would be used as any other WFS with some gain in an integrator loop.

\subsection{Model training}

For training Tokyo Drift, simple DNN architectures were used consisting of a stack of 2D convolutional layers \cite{lecun1989backpropagation} followed by multiple dense layers before the output.  ReLU activation \cite{nair2010rectified} was used in-between layers, with no additional activation after the output layers.  While more complex architectures may yield some improvements to performance, experience on similar problems has suggested that the gains from this strategy appear to be minor compared to the importance of carefully specifying the problem and crafting the datasets.  

Simulations were generated via {\ttfamily HCIPy} \cite{HCIPy}, using a custom module: {\ttfamily TelescopeSim}, built for use in machine learning tasks, with the ability to simulate filter bandwidth, detector and photon noise, deformable mirrors, Zernike aberrations in the pupil-plane, etc. \ref{telsim}.  Specifics of the NIRC2 instrument from work on 2F2F were used, including precise telescope and instrument pupil functions, filter specifics, plate scales, and details of the vortex coronagraph which were validated against on-sky operation.  Later iterations used similar details of the SCExAO VAMPIRES instrument \ref{ack_miles}.

For a supervised approach to training a DNN FPWFS using a single observation (i.e. with an asymmetric pupil \cite{kuhn2022small}), one simple approach is using the focal-plane image as the input, and as the target for regression, using the specific, randomly drawn Zernike errors imposed on the pupil-plane to generate them.  The loss function which proved most effective for these models is mean-absolute error (MAE).  Inputs and outputs were then stacked into batches (with typical batch size of 128).  Therefore, for a {\ttfamily 64x64} pixel focal plane image, predicting 35 Zernikes, the input batch would have the shape {\ttfamily (128, 64, 64, 1)}  (where the first dimension is batch size, and the final dimension indicates a single convolutional input channel), and the target batch shape would be {\ttfamily (128, 35)}.  However, to utilize the phase diversity offered by the F\&F technique, a more complex model was needed.

\begin{figure}
    \centering
    \includegraphics[scale=.35]{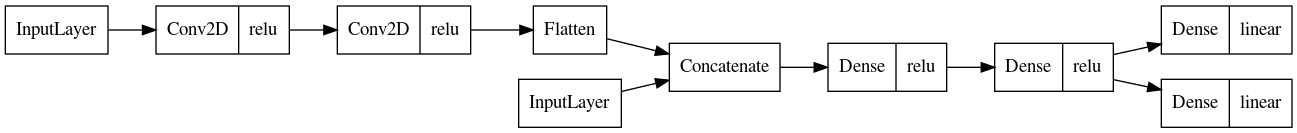}
    \caption{Sample Tokyo Drift model architecture utilizing the previous actuation as a separate input from the images.  This input is concatenated  with the outputs of the convolution stack.  Afterwards, a set of dense layers are utilized to fuse this information.  Finally, the output is broken up into the magnitude and sign of the errors with corresponding loss functions.}
    \label{fig:td_model}
\end{figure}

The F\&F algorithm requires two input images: the most recent and the one preceding it, as well as the differential deformable mirror actuation used between these frames.  As 2D convolutional layers were used for the initial layers of simpler models, adding the second image was simply accomplished by stacking it as an additional convolutional channel.  Adding the qualitatively different input vector of the intervening actuation meant this new model needed transition to multiple-inputs.  A schematic of how this was accomplished is illustrated in Figure \ref{fig:td_model} where the actuation vector input is fused with the output of the image convolution stack.  Therefore, with a batchsize of 128, the input now had a structure of {\ttfamily (128, 64, 64, 2)} for the input images, and {\ttfamily (128, 35} for a 35 Zernike deformable mirror actuation between frames.

\begin{figure}
    \centering
    \includegraphics[scale=0.3]{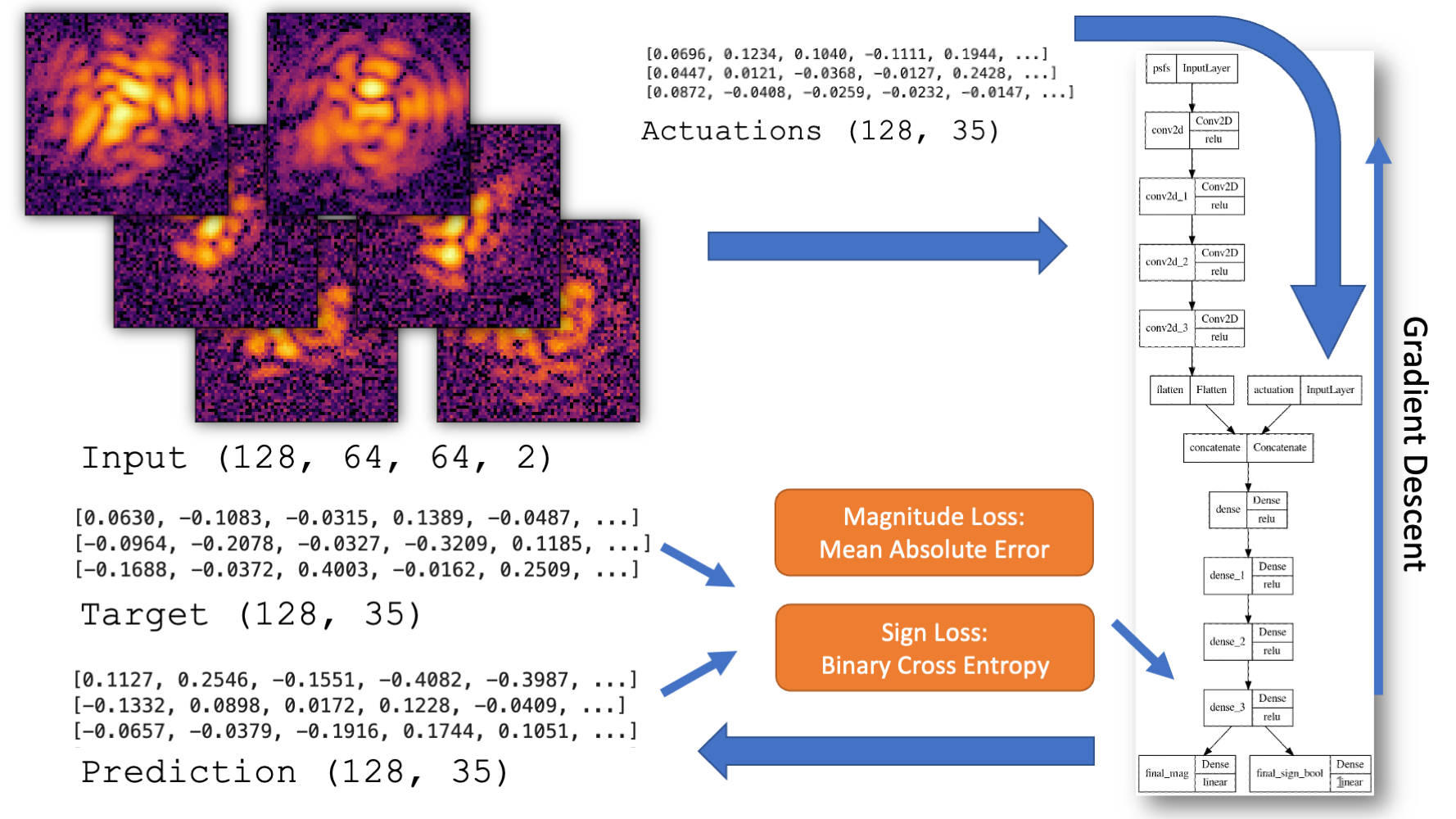}
    \caption{Schematic of the multi-input, multi-output Tokyo Drift architecture.  The model is trained on simulated data of the optical system, with inputs of two-image stacks, along with their intervening actuations and output of the predicted, best Zernike modes for correction.  The outputs are split into the magnitude of the Zernikes using MAE loss, and their sign using BCE.}
    \label{fig:td_arch}
\end{figure}

Further, it was recognized that while even modes were encoded ambiguously in the focal-plane, it was only the signs of these features which were unknown.  A DNN that is trained to regress on an single image of an even Zernike-mode ends up learning to predict zeros in order to minimize error while predicting odd modes correctly.  However, the structure in the image-plane which encodes the magnitude of these even aberrations should still be present to learn.  It was intuited that splitting up the prediction target into magnitude and sign should be useful, so each could be separately back-propagated with their own loss functions.  This meant the convolutional stack could focus on extracting the features encoding magnitudes, and the dense stack responsible for fusing in the actuation information could learn to correctly predict the signs as a separate task.  Moving to a multi-output model, the target batches were shaped {\ttfamily (128, 35)} for the magnitudes encoded as floating point numbers utilizing MAE loss, and {\ttfamily (128, 35)} for the signs of those targets, encoded as floats representing Boolean values and utilizing Binary Cross Entropy (BCE) loss.  A schematic of the full flow of inputs, outputs, and losses are shown in Figure  \ref{fig:td_arch}.

Constructing the distributions of errors for training this model was quite a bit more complex than previous, single-image models where it is only necessary to select a reasonable distribution of initial errors for each sample.  With this F\&F analog, a representative actuation for the consequent image also had to be constructed.  The final strategy employed was as follows:
\begin{enumerate}
    \item For the initial frame, draw random errors as before
    \item To construct a reasonable actuation, assume the FPWFS from first step is only somewhat accurate, and is running in a loop with some variable gain.  Therefore:
    \begin{enumerate}
        \item Take the true error of the first frame that would provide perfect correction and scale it by a random gain (for instance $[.01 - .99]$).  In fact, it proved important to scale beyond an ideal correction to simulate over-correction and subsequent errors flipping sign, so in the end, a simulated gain of $[0.01-1.99]$ was used.
        \item Add some extra, independent, random errors to simulate inaccurate sensing
    \end{enumerate}
    \item Finally, in order for the model to avoid simply propagating the signs from the input, and to provide a balanced training set for the BCE loss, the signs of the actuation used for the final frame were randomly flipped.  This simulated the FPWFS's blindness to sign in its first estimation.
\end{enumerate}

Several generations of models were produced in this fashion, necessitating large hyper-parameter sweeps over DNN architecture details, learning-rate, etc.  Bayesian hyperparamater optimization \cite{snoek2012practical} and Hyperband early stopping \cite{li2017hyperband} were used to search the large parameter spaces efficiently.  After which, the most successful models were trained to convergence.  
Along the way, these tests verified that the multi-output, multi-loss formulation did indeed provide a significant advantage, with a single output regression failing to learn much at all during training.  Additionally, after the initial architecture search was complete, further hyperparameter sweeps proved unnecessary for training TD models on completely different optical systems.  Simply generating a new dataset and retraining a model specific to those systems appears to be effective, allowing this technique to be quickly applied to different telescopes and their specific instrument configurations.

\subsection{Simulation results}

\begin{figure}
    \centering
    \includegraphics[scale=.4]{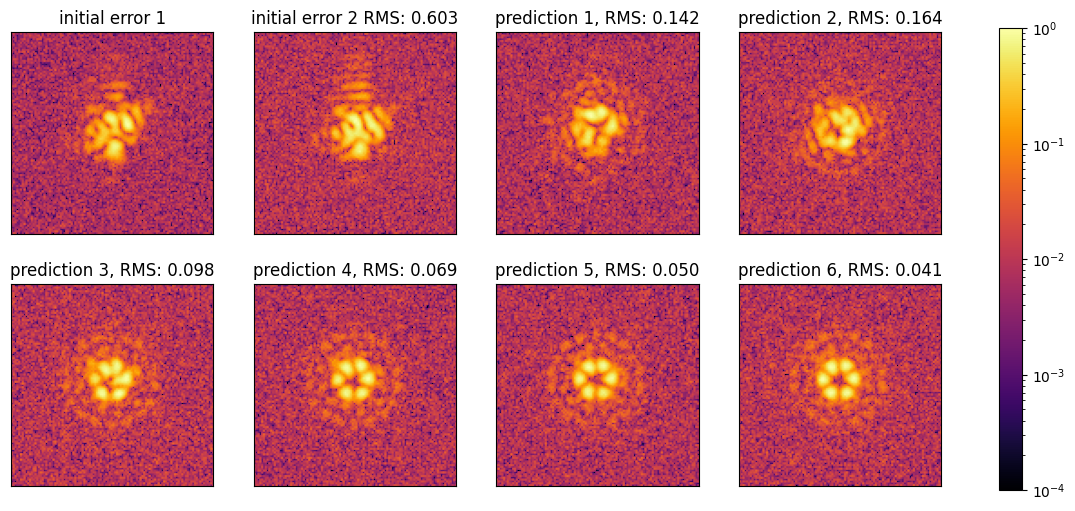}
    \caption{Example of iterative correction in an AO loop on a simulation of NIRC2 vortex coronagraph.  The first two frames are created via the training set generator.  These PSFs, plus the actuation between them, are fed into the trained DNN to produce the first prediction (third frame).  The process then repeats several more times until it converges to minimal wavefront error on the coronagraph.}
    \label{fig:td_coro_iter}
\end{figure}

The performance of this model was evaluated by simulating an adaptive-optics loop, where the initial state was created with the same dataset generation function as training.  Using the TD model then next predicted correction was applied with a gain of .6.  This process continued from there, iteratively feeding the predicted actuations and resulting observations back into the model (Figure \ref{fig:td_coro_iter}).

\begin{figure}
    \centering
    \includegraphics[scale=0.9]{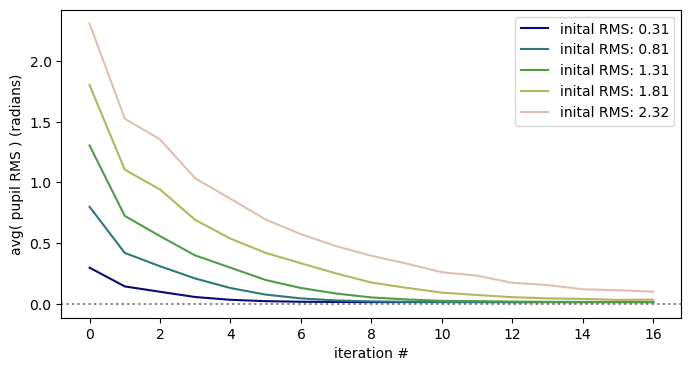}
    \caption{1000 random initial states with a wide range of initial errors were selected, then each was corrected over 16 iterations. The runs were grouped by initial RMS error and the results averaged here.}
    \label{fig:td_sim_coro_trials}
\end{figure}

The resulting model seems to be capable of correcting aberrations with surprisingly high initial RMS errors (Figure \ref{fig:td_sim_coro_trials}).  Typically, focal-plane sensing techniques are limited to around 1 radian RMS error before being unable to converge.  While errors this large should rarely be encountered in a stable coronagraphic instrument with high-quality traditional WFS, it is still promising to see that the technique has capabilities beyond what has previously been shown.

\subsection{Bench results}

Testing with real optical systems is currently underway on two fronts.  First, a simple adaptive-optics test-bed was built in the Institute for Astronomy - Hilo's optical lab.  This consists of a fiber-fed laser, which is collimated, the pupil then re-imaged on a 12x12 Boston Micromachines DM, and then imaged on a detector (\ref{ack_barnac}).  Aberrations are then introduced via the DM and sensed with a TD model trained on simulated data.  While there have been many difficulties with this setup -- being unstable optically, a stuck actuator and other issues with the DM, and the model not being trained on a very precise simulation of the system as it changes -- the Tokyo Drift model still performs well, converging with large initial input errors.

\begin{figure}
    \centering
    \includegraphics[scale=0.6]{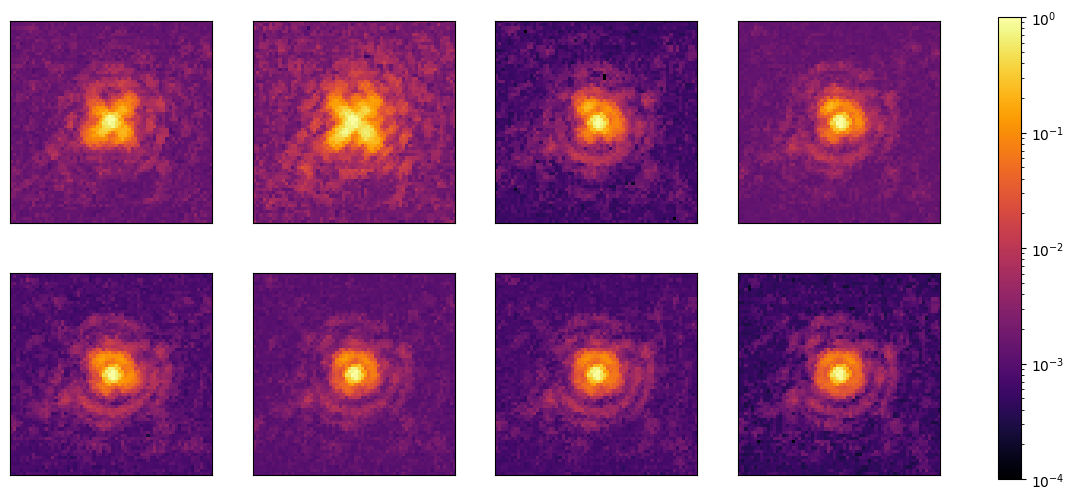}
    \caption{Data from a recent Tokyo Drift bench-test run on SCExAO's VAMPIRES instrument in the F760 filter without a coronagraph.  A model was trained to correct 10 Zernike-modes, random initial errors were chosen and set on the DM and iteratively corrected using only this technique.}
    \label{fig:td_vampires_bench}
\end{figure}

More recently, we have been doing daytime testing using the VAMPIRES instrument \cite{norris2015vampires, lucas_visible-light_2022} on Subaru.  The first model was trained without a coronagraph and with only 10 Zernikes (Figure \ref{fig:td_vampires_bench}), with the intention of scaling up the difficulty by adding higher order aberrations and utilizing the recently installed vector vortex coronagraph.  Again, with many small sim-to-real barriers to overcome, the Tokyo Drift model has shown successful convergence.  Both of these optical systems demonstrate that the success of the TD approach is viable for real optical systems.

\section{Conclusions}
We have presented two approaches to coronagraphic low-order wavefront control using a sequential method based on the Fast and Furious algorithm of Keller and Korkiakoski.  

The first, ``2 Fast 2 Furious'' is an extension of the linear methodology of F\&F to Lyot-style coronagraphs.  The simple implementation and advantageous properties of F\&F are preserved, though as before, a symmetric optical system is required, precluding usage with coronagraphic architectures like vortex phase masks.  Simulations show similar convergence properties of the original F\&F.  Bench testing indicates the algorithm works in principle, though further work is needed to demonstrate correction of multiple low-order modes.

The second approach, ``Tokyo Drift,'' uses a deep neural network trained on simulated images of the coronagraphic system with aberrations, but employs the same inputs and outputs as F\&F.  This gives it the ability to correct asymmetric optical systems, as well as outside the weak-aberration limit.  The drawback of this approach is the time it takes to train the network, though this is a one-time sunk cost.  Simulations show excellent performance with a vortex coronagraph, and lab testing shows very good convergence, albeit in a system without a coronagraph.  Bench tests with coronagraphs in place are now commencing.

It is intriguing to consider extending TD to dig dark holes in order to further enhance contrast.  One possibility for achieving this would involve pre-calculating Zernike patterns that create dark holes in a perfectly corrected system, and then generating new simulated datasets with perturbations, with the target of converging on these patterns.  In this way, the problem remains a supervised one, and so trivial to adapt the current training process.  Another possibility, is to reframe the problem using reinforcement learning (using the TD neural network as the ``policy,'' in reinforcement learning nomenclature).  The advantage is that the target for prediction does not need to be differentiable.  Instead, one could directly use measurements of the quality of the dark hole, e.g. minimizing {\ttfamily (flux in the desired dark spot location) / (central PSF brightness)}.  This has the advantages of not needing to guess at useful distributions of errors in the training set, learning better the model's own prediction errors, and the ability to use with real optical systems instead of simulations.  

Future work will focus on validation on coronagraphic testbeds.  Keck/NIRC2 finished a major detector upgrade and VAMPIRES on Subaru recently had both Lyot coronagraphs \cite{lucas_visible-light_2022} and a vortex coronagraph (Doelman et al. in prep) installed, which will facilitate further testing and validation.

\section{Code, Data, and Materials}

\label{telsim} Framework for simulating telescope systems with HCIPy, formulated for running machine learning experiments https://github.com/icunnyngham/TelescopeSim

%\subsection{References}

%For books\cite{Lamport94,Alred03,Goossens97}, the listing includes the list of authors, book title, publisher, city, page or chapter numbers, and year of publication.  A reference to a journal article\cite{Metropolis53} includes the author list, title of the article (in quotes), journal name (in italics, properly abbreviated), volume number (in bold), inclusive page numbers, and year.  By convention\cite{Lamport94}, article titles are capitalized as described in Sec.~\ref{sec:title}.  A reference to a proceedings paper or a chapter in an edited book\cite{Gull89a} includes the author list, title of the article (in quotes), volume or series title (in italics), volume number (in bold), if applicable, inclusive page numbers, publisher, city, and year. For websites\cite{Lees-Miller-LaTeX-course-1} the listing includes the list of authors, title of the article (in quotes), website name, article date, website address either enclosed in chevron symbols ('\(<\)' and '\(>\)'),  underlined or linked, and the date the website was accessed. 

%If you use this formatting, your references will link your manuscript to other research papers that are in the CrossRef system. Exact punctuation is required for the automated linking to be successful. 

%Citations to the references are made using superscript numerals, as demonstrated in the above paragraph.  One may also directly refer to a reference within the text, e.g., ``as shown in Ref.~\cite{Metropolis53} ...''

\acknowledgments 

\label{ack_miles}Thanks to Miles Lucas for providing precise optical system details for SCExAO's VAMPIRES instrument as well as extensive help running tests on-bench.

\label{mana}The technical support and advanced computing resources from University of Hawaii Information Technology Services – Cyberinfrastructure, funded in part by the National Science Foundation CC* awards \# 2201428 and \# 2232862 are gratefully acknowledged.

\label{ack_barnac}Thanks to Christoph Baranec for the use of his deformable mirror and other important optical components for the IfA AO bench.  Thanks to James Ou for help with interfacing with the DM.  Thanks to Charles-Antoine Claveau and Aidan Walk for help interfacing with the detector.

% References

\printbibliography %Prints bibliography
\end{document}